# Electrical Readout Strategies of GFET Biosensors for Real-World Requirements

Michael Geiwitz[a], Owen Rivers Page[b], Marina E. Nichols[a], Tio Marello[a,b], Catherine Hoar[c], Deji Akinwande[d], Michelle M. Meyer[b], Kenneth S. Burch[a]

[a] *Department of Physics, Boston College, Chestnut Hill, MA 02467, USA*

[b] *Department of Biology, Boston College, Chestnut Hill, MA 02467, USA*

[c] *Department of Engineering, Boston College, Chestnut Hill, MA 02467, USA*

[d] *Department of Biomedical Engineering, The University of Texas at Austin, Austin, TX 78758, USA*

*Corresponding Author - *Email*: ks.burch@bc.edu (K.S. Burch)

## Abstract

Graphene Field-Effect Transistors (GFETs) are increasingly employed as biochemical sensors due to their exceptional electronic properties, surface sensitivity, and potential for miniaturization. A critical challenge in deploying GFETs is determining the optimal electrical readout strategy. GFETs are typically operated with either of two modalities: one measuring current in real time (amperometric) and the other monitoring the change in voltage for charge neutrality (potential potentiometric). Here, we undertake a systematic study of the two modalities to determine their relative advantages/disadvantages towards guiding the future use of GFETs in sensing. We focus on viral proteins in wastewater, given the matrix's complexity and the growing interest in the field of wastewater surveillance. Our results show that transconductance offers far superior limits of detection (LOD) but suffers from limited reproducibility, a narrower dynamic range, and is ineffective for some viral proteins. In comparison, we find that Dirac point tracking offers higher reproducibility and superior robustness, but at a higher LOD. Interestingly, both techniques exhibit similar sensitivity, highlighting the importance of the aptamers employed. Systematic experiments also help explain differences in dynamic range and limited functionality in detecting some proteins, resulting from hidden electrophoresis, and shifting the high transconductance point away from the active region. Thus, our findings provide crucial considerations for designing and operating resilient graphene biosensors suitable for real-time pathogen monitoring in environmental scenarios.

## 1. Introduction

Graphene Field-Effect Transistors (GFETs) have emerged as a versatile platform for biosensing due to their exceptional electrical properties, biocompatibility, and high surface-to-volume ratio (Geim and Novoselov, 2007; Geiwitz et al., 2024; Hu et al., 2023; Krishnan et al., 2023). These characteristics make them particularly well-suited for the detection of biomolecular analytes, including viruses, for which sensitivity and specificity are paramount (Kumar et al., 2023; Sengupta and Hussain, 2021; Seo et al., 2020). To date, efforts have focused on proof-of-concept or on enhancing GFET robustness to biofouling in complex media (Gao, Zhaoli and Xia, Han and Zauberman, Jonathan and Tomaiuolo, Maurizio and Ping, Jinglei and Zhang et al., 2018; Lozano-Chamizo et al., 2024; Sengupta and Hussain, 2021; B. Sun et al., 2025). Nonetheless, an overlooked aspect in optimizing GFET-based biosensors is determining the most effective electrical readout method for a given application (Krishnan et al., 2023; Ono et al., 2024).

Among the most used electrical readout strategies in GFET biosensing are Dirac point voltage shift measurements and transconductance analysis, each of which interrogates different aspects of the device's response to surface-bound analytes. These two options arise from differences between FET technologies based on graphene versus and those based on standard semiconductors. Graphene is a semimetal in which the conduction and valence bands meet at a single point in momentum space, called the Dirac point. At this point, the density of electronic states vanishes, and the Fermi level lies at the charge neutrality point (CNP) (Castro Neto et al., 2009; Das Sarma et al., 2011). In a GFET, this manifests as the gate voltage $(V_G)$ at which the drain-source resistance $(R_D)$ reaches a maximum (Ohno et al., 2010). Thus, it is typically assumed that when a target analyte binds to the probe on the graphene surface, it induces a shift in the carrier concentration, effectively doping the graphene and shifting the Dirac point to a new gate voltage (Haslam et al., 2018; Kim et al., 2013). This is often used as a label-free indicator of molecular binding events (Kim et al., 2013; Ohno et al., 2010). Furthermore, since the transferred charge should be proportional to the total charge from the bound targets, the Dirac point shift can be calibrated to determine the target concentration(Haslam et al., 2018).

In contrast to Dirac voltage tracking, like standard FET biosensors, GFETs have been employed by monitoring the source-drain current $(I_{SD})$ for fixed source-drain $(V_{SD})$ and gate voltages. This is typically performed at the point of maximum transconductance ($\frac{d\sigma}{dV_G}$) (Giambra et al., 2019; Meric et al., 2013; Schwierz, 2013; Szunerits et al., 2024). Here, one similarly relies on the shift in the Dirac point upon analyte binding to the probe on the surface, which produces a change in $I_{SD}$ (Béraud et al., 2021; Szunerits et al., 2024; Ushiba et al., 2023). However, it should be noted that such changes can also occur due to reduced carrier mobility (Balandin, 2011; Giambra et al., 2019; Meric et al., 2013; Schwierz, 2010). Since $\Delta I_{SD}$ is measured at the initial point of maximum transconductance, it should have much higher sensitivity than shifts in the Dirac point, where $\frac{d\sigma}{dV_G} \sim 0$ (Béraud et al., 2021; Fuhr et al., 2023; Giambra et al., 2019; Rodrigues et al., 2022). However, this increased sensitivity may come at the expense of reproducibility (Ushiba et al., 2023). Furthermore, since the applied $V_g$ is typically fixed, transconductance measurements will continuously move to regions of lower transconductance and thus are likely to have a lower dynamic range(Fuhr et al., 2023). Both electrical readout methods have been employed by various groups in previous studies (Béraud et al., 2021; Giambra et al., 2019; Kumar et al., 2022, 2023), but so far no systematic tests have established the conditions under which each method is more effective.

Understanding the utility of these two readout modalities is critical for optimizing GFET-based sensors, particularly in applications where sensitivity, selectivity, and reproducibility must be balanced against sample complexity and operational constraints (Krishnan et al., 2023; Szunerits et al., 2024). This question has gained urgency with the widespread deployment of GFETs for numerous sensing applications, ranging from pH levels in sweat to opioids and viruses in complex media such as wastewater (Geiwitz et al., 2024; Krishnan et al., 2023; Lerner et al., 2014; Wang et al., 2022). Here, we focus on detection of target analytes of intermediate size in a complex medium: viral proteins in wastewater (Geiwitz et al., 2024). Furthermore, both electrical readout modalities could be helpful for different applications of wastewater surveillance. For example, the increased sensitivity achieved by monitoring transconductance may more appropriate for early detection of emerging infectious diseases; during the early stages of an outbreak, establishing presence or absence of the associated pathogen in wastewater is more crucial than measuring its concentration or optimizing selectivity (Bivins et al., 2020). On the other hand, approaches that measure shifts in the Dirac point may be better suited for longitudinal wastewater surveillance aimed at tracking disease prevalence, since such studies require comparison of pathogen concentrations and optimal measurement accuracy over time (Hughes et al., 2023). In both cases, the use of low-cost, field-deployable sensors for point-of-need pathogen tracking would offer a revolutionary advance in expanding access to wastewater surveillance to support infectious disease epidemiology and prevention (Geiwitz et al., 2024; Hughes et al., 2023; Kumar et al., 2022).

We note that another key challenge in conducting our current study is the need for highly reproducible fabrication (Ramoso et al., 2025; Sengupta and Hussain, 2021; Soikkeli et al., 2023). With this in mind, we recently developed a method to create GFET devices at wafer scale that can detect four different analytes in parallel (Geiwitz et al., 2024; Kumar et al., 2020a, 2022). This Graphene Electronic Multiplexed Sensor (GEMS; see Fig. 1) is highly reproducible across dozens of fabrication and testing iterations, with consistent initial Dirac points(Geiwitz et al., 2024). GEMS has also proved highly robust in wastewater matrices with LODs lower than those achieved with liquid chromatography-mass spectrometry (LC-MS), a finding consistent with broader trends in biosensor-based wastewater surveillance platforms(Azubuike et al., 2022; Geiwitz et al., 2024; Kumar et al., 2021; Mao et al., 2021). However, our previous efforts employed only the modality of monitoring changes in the Dirac point (Geiwitz et al., 2024; Kumar et al., 2020b, 2022).

Thus, to determine the viability of GEMS for real-time monitoring and understand the relative benefits and drawbacks of the two readout strategies, we systematically compared transconductance and Dirac voltage shift modalities for aptamer-functionalized GFETs for viral protein detection in wastewater. Specifically, we elected to focus on detection of SARS-CoV-2, Influenza, and Respiratory Syncytial Virus (RSV), three common respiratory viruses which are detectable in wastewater and are current targets of many wastewater surveillance programs, including the U.S. Centers for Disease Control and Prevention (CDC) National Wastewater Surveillance System (NWSS) (Adams et al., 2024; CDC, n.d.; Jones et al., 2020; Lowry et al., 2023; Medema et al., 2020; Naughton et al., 2023). For each analyte and readout modality, we assess GEMS performance in terms of sensitivity, reproducibility, and noise tolerance.

The findings in this study offer a practical framework for selecting the optimal GFET readout modality. We compare results from a top-gate electrode with those from a coplanar side gate, demonstrating the latter's advantages (lower LOD and enhanced reproducibility). Advantages and disadvantages of both Dirac point tracking and transconductance measurements are also outlined. Of note is the case of charged analytes being attracted or repelled by the gate during transconductance measurements. These results advance the development of robust, field-deployable biosensors for real-time epidemiological surveillance in complex sample matrices, such as wastewater.

## 2. Materials and methods

### 2.1 Materials

Recombinant viral proteins were obtained from a commercial supplier (Acro Biosystems) and reconstituted in ultrapure water according to the manufacturer's protocols. Serial dilutions of each antigen were prepared in two matrices: (1) diluted, sterile phosphate-buffered saline (0.01x PBS, pH 7.4), and (2) diluted municipal wastewater (10:1 with DI water), based on our previous work (Geiwitz et al., 2024).

Raw wastewater samples were collected from the influent of a local wastewater treatment plant and filtered through 0.3 µm syringe filters to remove debris. Before the addition of exogenous viral proteins (i.e., spiking the samples), all samples were equilibrated to room temperature and vortexed to ensure homogeneity. Target concentrations ranged from 1 ag/mL to 10 µg/mL to cover both the environmentally relevant and low-abundance detection regimes. Wastewater control samples containing no exogenously-spiked analyte were used to assess background response and nonspecific interactions. Negative control samples of maximally applied concentrations off-target analytes were also used to test aptamer specificity.

### 2.2 Device Fabrication

As in our previous work (Geiwitz et al., 2024; Kumar et al., 2020b, 2022), GFETs were fabricated on standard degenerately doped p-type silicon wafers (resistivity < 0.005 Ω·cm). The source and drain electrodes for 44 individual chips are patterned concurrently on the wafer using a standard bi-layer photolithography process with a µMLA (Heidelberg Instruments) maskless system. After the bottom contacts are patterned and deposited, the wafer is sent to General Graphene, where a full wafer monolayer of graphene is transferred onto the bottom contacts. Once returned, residual transfer residues are removed chemically and thermally. A 3 nm layer of aluminum oxide is deposited to protect the graphene during the remainder of the fabrication process.

The wafer then undergoes another bi-layer photolithography process to create the graphene etching pattern. During the pattern development, the MF-319 not only develops the lithography pattern but also etches the exposed aluminum oxide beneath the photoresist layers. Oxygen plasma at 60W is applied for 60s to etch the graphene. Then, a 50 nm aluminum oxide passivation layer is deposited. Unlike in previous works, we now perform an additional round of photolithography to pattern the on-chip side gates, with the same thickness (5 nm Ti/20 nm Pt) as the source and drain electrodes. We have found that fabricating the side gates after graphene etching enhances reproducibility and lowers the measured initial Dirac point which likely results from preventing oxidation of the platinum during the oxygen plasma process (Angerstein-Kozlowska et al., 1973; Tan et al., 2022). The next step of fabrication uses a single-layer lithography process to open 10μm x 40μm windows to the graphene through the passivation layer and expose the contact pads for wire bonding. The wafer (Figure 1a) is diced into individual chips approximately 1 cm$^2$ in area (Figure 1b). The chips are then mounted with double-sided tape and wire-bonded to custom-made PCBs designed to fit into our proprietary measuring device.

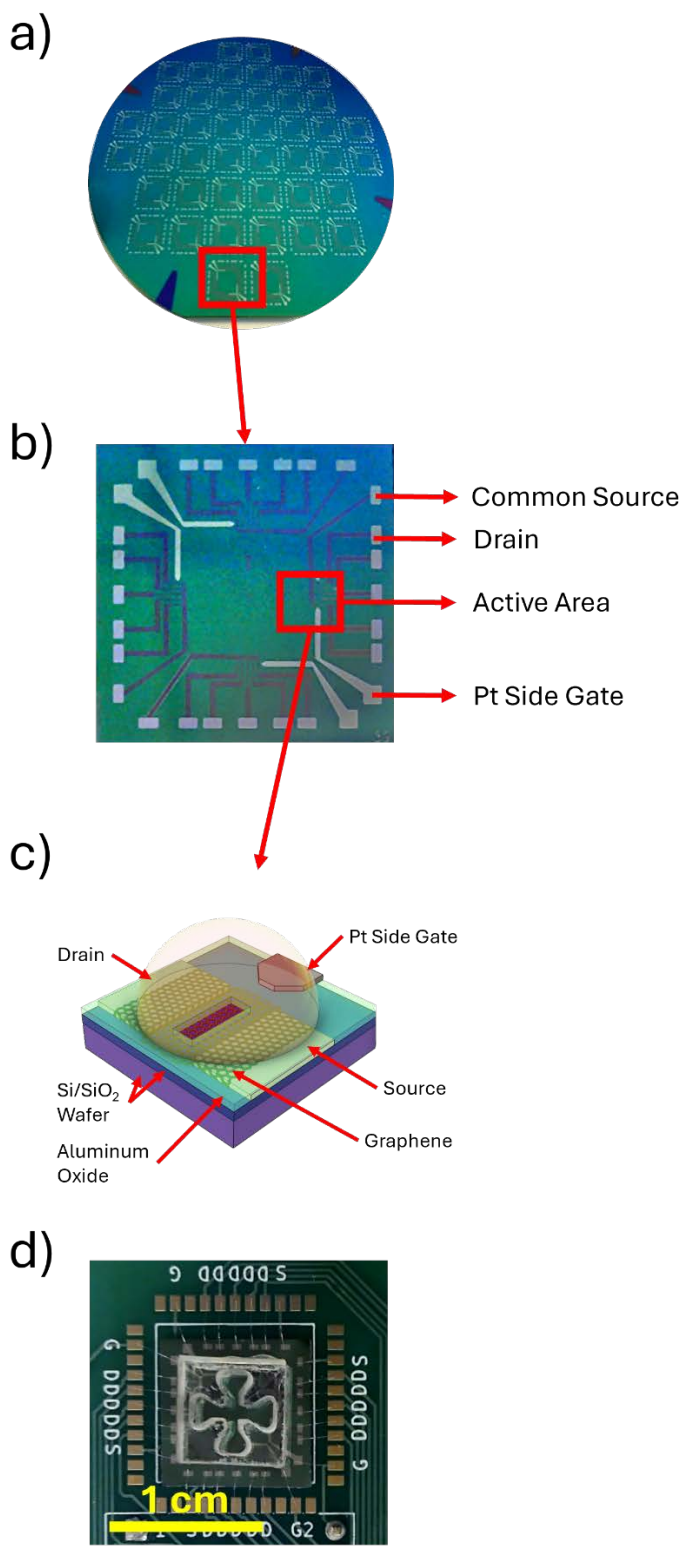

Figure 1. a) Wafer containing 44 GFET chips pre-dicing. b) Enlarged chip post-fabrication. c) Schematic of a single GFET device including on-chip side gate. d) Single chip mounted on custom PCB with applied acrylic well.

Unlike our previous efforts, we have made a few changes to the wells used for functionalization and testing. As in other reports, we used PDMS wells to enable separate functionalization of each GFET group. However, we found limited reproducibility and a long drying period before they became useful (Figure S1). In addition, our previous design required manual placement of the sample into wells functionalized with specific aptamers. Thus, to move towards real-time monitoring and enhance reproducibility, we switched to acrylic wells pre-attached to pressure-sensitive adhesive (FlexDym). These were then laser cut before being adhered to the GEMS chips. We found this dramatically improved reproducibility and allowed testing within a few hours of production. Furthermore, we employ a "clover"-shaped well, which enables us to individually functionalize each sensing area with a different probe (Figure 1c) and to apply the sample to all of them simultaneously. Specifically, each "leaf" provides sufficient surface tension of the liquid such that each probe is suspended in a single area (Figure S2). Each leaf contains five GFET devices and an on-chip side gate. A schematic of a single GFET device with source, drain, and gate electrodes, along with the graphene and aluminum oxide passivation layers, is shown in Figure 1c.

## 2.3 Surface Functionalization and Preparation

To impart target specificity, GFETs are functionalized with aptamer probes targeting three viral surface proteins: the SARS-CoV-2 spike protein (spike), influenza A hemagglutinin (HA), and the respiratory syncytial virus (RSV) glycoprotein. These same aptamers have already successfully enabled selective, multiplexed protein detection in wastewater using GEMS. Aptamers were pre-attached to 1-Pyrenebutyric acid N-hydroxysuccinimide ester (PBASE) molecules (Supplemental S3) to prevent graphene exposure to DMF, which we found dramatically enhances reproducibility and LOD (Geiwitz et al., 2024). The PBASE molecules attached to the aptamers adsorb onto the graphene surface via non-covalent π–π stacking, forming a stable linker layer without disrupting graphene's electronic structure (Krishnan et al., 2023; Mishyn et al., 2022).

## 2.4 Electrical Measurements

All electrical measurements were made with our proprietary measurement platform, built from off-the-shelf components and programmed in Python. To determine the gate voltage with the maximum transconductance, we first measure the conductance as a function of gate voltage after aptamer attachment (Figure 2). The peak transconductance point is found by numerically calculating the first derivative of conductance with respect to voltage. This derivative is not the true transconductance, which differs only by a scaling factor of $V_{DS}$ (Lee et al., 2012; Pacheco-Sanchez et al., 2020; Ryzhii et al., 2011),

$$g_m = V_{DS}\left(\frac{\delta G}{\delta V_{gs}}\right) \quad (1)$$

but clearly shows the voltage at which the transconductance reaches an absolute maximum in both the hole and electron regimes of the transport curve. We select the peak in the hole regime (Figure 2a) because, as higher concentrations of analyte are added, the transport curve shifts to higher $V_G$, resulting in a slower reduction of $g_m$ with concentration. We then set the static gate voltage in our measurement software to the voltage at which this peak occurs.

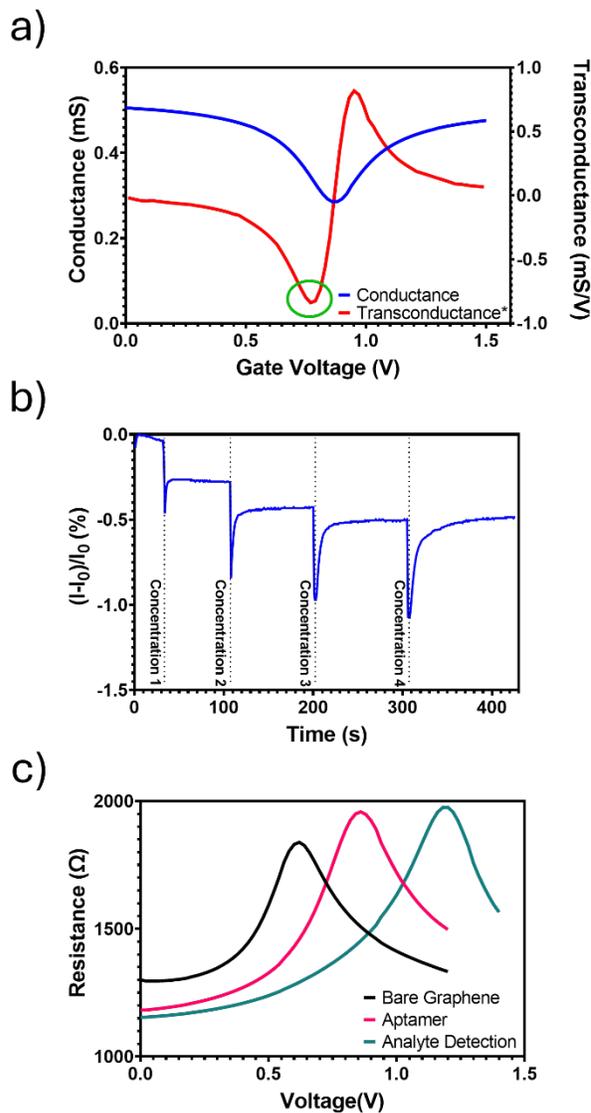

Figure 2. a) Measured conductance (blue) were found and the first derivative (*which is proportional to transconductance) plotted (red). The green circle shows the peak transconductance point in the hole regime of the graphene. b) Example plot of a typical transconductance measurement. Sudden drops in current indicate the points at which different concentrations of target analyte are added to the device well. c) Example Dirac point voltage tracking plot showing the initial Dirac point of unfunctionalized, bare graphene, the shifted transfer curve after aptamer attachment, and a further shift after exposure to a given target analyte.

To perform real-time measurements, we first add wastewater 10x diluted in PBS to the well as our ionic gating solution. With the gate voltage set and active, we begin to measure the current as a function of time. We wait for the ions in the solution to settle, indicated by the current reaching a stable point, then begin adding negative controls or target analytes. This waiting period is also employed between subsequent analyte injections. Different and increasing concentrations of wastewater spiked with the target analyte are added to the GFET well until no significant drop in current is observed (Figure 2b). An example of a Dirac point voltage tracking is shown for comparison in Figure 2c. In short, the initial Dirac point voltage of the bare graphene is found as an indicator of graphene quality. After aptamer functionalization, the Dirac point is again measured and it is this point that serves as the reference where all future Dirac point voltage shifts, such as the analyte detection curve, are measured.

To determine the limit of detection (LOD), we employed the stabilized $I_{SD}$ signal from the negative control as the baseline. Next, the percentage in the current drop is measured after applying a specific target analyte concentration and plotted as a function of concentration. These are then fit using Hill's equation, from which the dynamic range and LOD can be determined from the fitting factors (Borisov et al., 2000; Motulsky and Christopoulos, 2004; Weiss, 1997). We account for differences in the injected and the actual analyte concentration in the GFET well, resulting from adding to the current volume and concentration of the liquid. An example of injected versus actual concentrations during a single measurement is shown in Table 1. The actual concentration in the well after a new concentration is injected is given by:

$$C_{actual} = \frac{C_0 V_0 + C_1 V_1}{V_0 + V_1} \tag{2}$$

where $C_0$ and $V_0$ are the concentration and volume presently in the well and $C_1$ and $V_1$ are the concentration and volume of the added sample.

**Table 1.** *Injected exogenous target protein concentration and actual concentration of target protein in the GFET well. Each well started with 60 µL of diluted wastewater with no exogenous target proteins spiked in. The left column shows the concentration of the exogenous protein target spiked into the diluted wastewater and added to the well; the right column shows the running total measured concentration after each injection.*

| Injected Concentration (10 µL) | Actual Concentration in Well |
|---|---|
| 1 ag/mL | 0.2 ag/mL |
| 1 fg/mL | 0.166 fg/mL |
| 1 pg/mL | 0.143 pg/mL |
| 1 ng/mL | 0.125 ng/mL |
| 1 µg/mL | 0.111 µg/mL |

Another critical performance metric for any biochemical assay, biosensor, or dose–response model is the linear dynamic range (LDR) because it defines the concentration interval over which the measured signal is proportional to the analyte level (Calvert, 1990; Liu et al., 2014; Prabowo et al., 2021; Purohit et al., 2020). A broad LDR enables accurate quantification across multiple orders of magnitude without requiring sample dilution or specialized calibration strategies (Prabowo et al., 2021; Purohit et al., 2020). In contrast, a narrow LDR restricts quantifiable measurements to a small concentration window and may introduce significant uncertainty or nonlinearity-based bias when analyte levels fall outside this region.

In sigmoidal systems described by the Hill equation, the LDR corresponds to the central transition region of the curve, where the slope is approximately linear and the log-transformed dose–response relationship is well approximated by a line. In the context of the Hill equation (Motulsky and Christopoulos, 2004; Weiss, 1997),

$$\theta = \frac{[L]^n}{K_d^n + [L]^n}, \tag{3}$$

where $\theta$ is the fractional response, $[L]$ is the ligand concentration, $K_d$ is the apparent dissociation constant, $n$ is the Hill coefficient, and the slope of the linear region in the Hill plot is proportional to $n$. The Hill coefficient thus governs the steepness of the sigmoidal dose–response curve and inversely determines the width of the linear dynamic range. This region is essential for analytical applications because parameter estimation (e.g., $K_d$ or sensitivity) is most reliable where the slope is linear. In biosensing contexts, the LDR determines the useful detection range of the sensor and dictates whether the device can accommodate the expected physiological or environmental concentration variations (Munje et al., 2015; Sales da Rocha et al., 2025).

To quantify the LDR, if the linear range is operationally defined as the concentration interval between 10% and 90% response, the corresponding ratio of ligand concentrations is given by (Borisov et al., 2000; Motulsky and Christopoulos, 2004):

$$\frac{[L]_{90}}{[L]_{10}} = \left(\frac{0.9/(1-0.9)}{0.1/(1-0.1)}\right)^{1/n} = 81^{1/n}. \tag{4}$$

Taking the logarithm gives the approximate width of the linear dynamic range on a logarithmic concentration scale:

$$\log_{10}\left(\frac{[L]_{90}}{[L]_{10}}\right) \approx \frac{1.91}{n}. \tag{5}$$

Hence, the linear dynamic range of a Hill-type response decreases inversely with the Hill coefficient. systems with larger *n* exhibit narrower concentration window over which the response is linear.

# 3. Results and Discussion

## 3.1 SARS-CoV-2 spike proteins and RSV glycoprotein

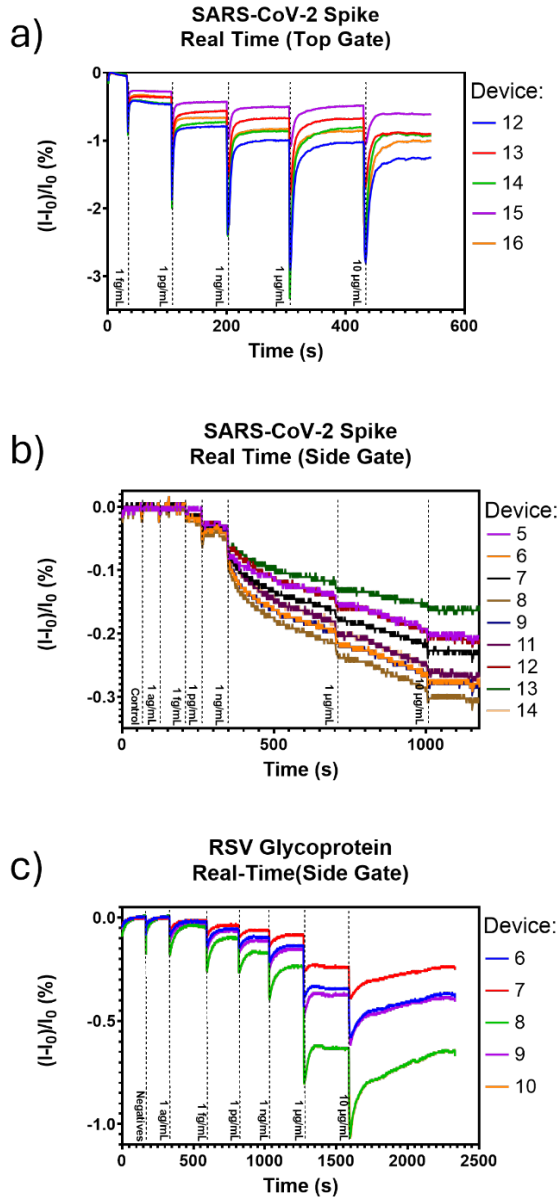

Figure 3. a) Real-time SARS-CoV-2 measured with platinum wire top gate. b) Real-time SARS-CoV-2 detection measured with platinum, co-planar side gate. c) Real-time current vs. time measurement for RSV glycoprotein in wastewater measured using the side gate.

### 3.1.1 Top gate vs. side gate

We note that many electric biosensing experiments have used top gates, in which a wire is inserted into the liquid. However, such approaches are not amenable to scale-up, and their reliability is unclear. Thus, we compared results from a platinum wire top-gate with those from our on-chip platinum side-gate. Initial experiments with the SARS-CoV-2 spike protein revealed distinct differences. Consistent with results from others (Kumar et al., 2023; Rainey et al., 2023; Xu et al., 2022), the top-gate measurements exhibited pronounced current dips or overshoot (Figure 3a) immediately after adding a new concentration, suggesting transient instability in the ionic double layer. These abrupt current decreases likely originate from physical and electrostatic disturbances (Xue et al., 2022; Yildiz et al., 2021): (1) pipette-induced turbulence disrupts the ion distribution at the graphene–electrolyte interface, temporarily reducing gate coupling efficiency, and (2) progressive submersion of the platinum electrode alters the effective gating area, changing the capacitive coupling to the channel. Both mechanisms contribute to transient current suppression and introduce artifacts that complicate the interpretation of binding-induced signals.

We then repeated the measurement using the side gate. Here, we found the dips in current after concentration additions were nearly absent (Figure 3b). This suggests that the top gate being further submerged while adding volume to the well may have a stronger effect than the ion disturbance at the graphene.

To determine whether the differences observed in the transconductance approach with the SARS-CoV-2 spike protein are more general, we measured the respiratory syncytial virus (RSV) glycoprotein. The RSV g-protein is much smaller (27 kDa (McLellan et al., 2013)) than the spike proteins (140 kDa (Zhu et al., 2021)), so the constant electric field supplied by the side gate may have different effects. Minor post-injection current dips were still observed (Figure 3c), implying residual ion redistribution, but these transients were less severe than with the top gate and stabilized rapidly.

### 3.1.2 Limits of detection

As outlined above, all the current changes were fit to Hill's equation (Figure 4a and 4b). The average LODs using the top gate for the spike proteins for the measured devices were 275 ± 10.3 ag/mL. This is several orders of magnitude lower than the LOD of $1.36 \times 10^5 \pm 1.2 \times 10^4$ ag/mL we reported in our previous work using Dirac point-shift measurements (Geiwitz et al., 2024). Interestingly, the LOD improved further when using the side gate (Figure 4a). With the side gate, we found the LOD to be an order of magnitude less, at 16 ± 0.25 ag/mL, than with the top gate. We thus attribute the superior stability of the side-gate configuration to its fixed geometric relationship with the graphene surface and its minimal hydrodynamic interference. As a result, the transconductance-derived signal in the side-gated devices exhibited a higher signal-to-noise ratio (SNR) improved baseline stability, and reduced hysteresis. Compared with our previous work, these results suggest that the transconductance approach can produce a much lower LOD (See Table 2).

To comprehensively compare the two approaches, we show in Figure 4a (4b) the concentration dependence for SARS-CoV-2 (RSV) for both transconductance and Dirac point methods. The measured LOD for RSV of 10.8 ± 1.2 ag/mL (Figure 4b) also represented a several-order-of-magnitude improvement over Dirac-based detection in our previous work ($1.76 \times 10^5 \pm 2.3 \times 10^4$ ag/mL), confirming the robustness of the differential transconductance readout under complex matrix conditions.

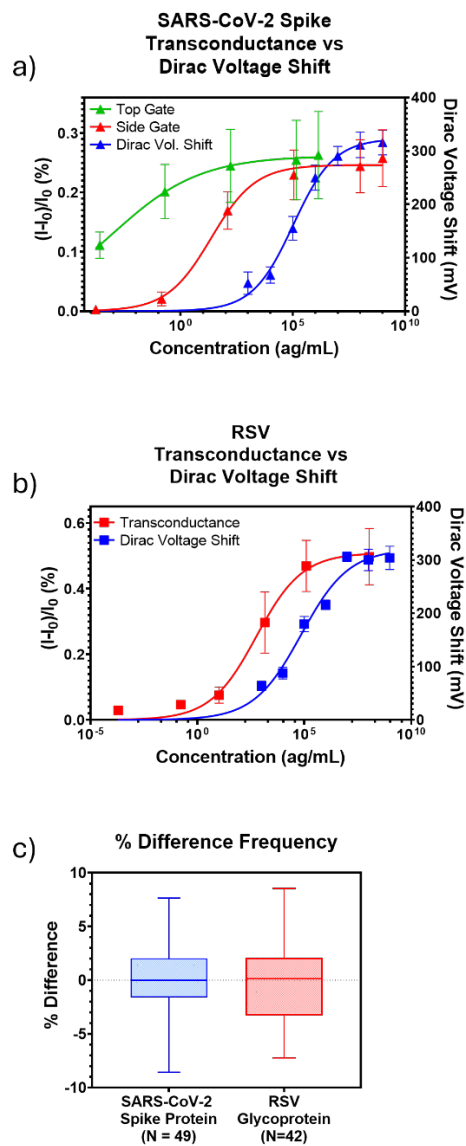

### 3.1.3 Linear Dynamic Range

When comparing both methods, we found that the LDR in the transconductance mode for SARS-CoV-2 spike proteins differed substantially from that in our previous work using the Dirac voltage shift. The Hill slope (Figure 4a – red) was calculated to be n = 0.50, giving an LDR of 3.8 orders of magnitude for the transconductance approach. With SARS-CoV-2 spike proteins in wastewater using Dirac point tracking (Figure 4a – blue), the Hill slope was n = 0.42, giving an LDR of 4.6 orders of magnitude, nearly a whole order of magnitude broader. Similarly, for the RSV g-protein using transconductance, the LDR was determined to be 4.5 orders of magnitude (n = 0.42), and with Dirac point tracking, the LDR is 4.9 orders of magnitude (n = 0.39).

### 3.1.4 Device Variability

Also of note is the variability between devices. The transconductance showed a high degree of difference in signal between devices as the concentration of analyte increased, as seen from the error bars in Figures 4a and 4b. The Dirac point measurements showed much less variability between the ten devices measured in that study. This variability in transconductance could be attributed to differences in the gate voltage corresponding to peak transconductance across devices. On our GEMS chip, one gate serves up to five GFETs, each with slight variability in peak transconductance voltage. Because of this, the voltage used during measurements is the average of the peak voltage for that set of

Figure 4. a) Real-time current vs. time measurement for SARS-CoV-2 in wastewater measured using the side gate. Sensitivity is improved using the side gate while the LDR is contracted when compared to Dirac Voltage Shift. b) Concentration dependence measurements of transconductance vs. Dirac voltage shift. Transconductance shows higher variability between devices as concentrations increase whereas Dirac voltage shifts are consistent. (blue – *Data from our previous work). c) Percentage difference between an individual device's peak transconductance voltage and the voltage set on the side gate

GFETs. Given the high sensitivity of devices near the peak transconductance voltage, a single GFET could experience a larger shift in its signal than its neighbors. Conversely, the derivative near the Dirac point is zero, so it is less sensitive to small changes and hence exhibits less variability (Ono et al., 2024; M. Sun et al., 2025). For example, Figure 4c shows a plot of the variability in the percentage difference between the actual peak transconductance voltage of an individual GFET and the set gate voltage used.

## 3.2 Influenza A Hemagglutinin Detection

Despite similar responses in Dirac Voltage mode, transconductance measurements targeting influenza A hemagglutinin (HA) showed a starkly different response when similarly compared. As shown in Figure 5a, in transconductance mode, HA binding produced negligible current changes until analyte concentration injected analyte concentration reached 10 ng/ml, nearly nine orders of magnitude higher than those at which the other proteins began to induce current drops. Furthermore, the percentage change in the current was an order of magnitude smaller. To ensure aptamer activity, we performed control experiments using the Dirac point protocol, which matched our previous experiments (Figure S4). Since the key difference in transconductance measurements is the applied gate (i.e., electric field) during target attachment, these results suggest the lack of early response arose from electrostatic and kinetic factors.

To see how electrostatics play a role, consider that the isoelectric point of HA (~5.0 (Kordyukova et al., 2019)) means that at physiological pH (~7.4), these proteins carry a net negative charge (Mapiour and Amira, 2023; Tokmakov et al., 2021). When a positive bias is applied to the gate, these negatively charged proteins experience an attraction to the gate electrode, effectively reducing their local concentration near the graphene surface (Figure 5c). This mechanism explains the observed insensitivity at low concentrations in the presence of an active gating field. In contrast, at higher analyte concentrations, diffusion-driven encounters between aptamer and target overcome this electrostatic barrier, leading to measurable current modulation. In the absence of an applied field, the analytes are free to move by Brownian motion to find and bind to the aptamer probes (Figures 5d and 5e). When compared to the other target analytes in this work, the RSV glycoprotein has an isoelectric point of ~9.0 (Chai et al., 2024; Voorzaat et al., 2024), and the SARS-CoV-2 spike protein is in the range of 6.0 – 8.0 (Hristova and Zhivkov, 2024). An isoelectric point greater than 7.4 means that the analytes will have a positive or near-neutral charge when suspended in a near-neutral pH medium. Thus, while the HA proteins are attracted to the gate, the RSV and spike proteins are repelled and pushed to the graphene surface. The range in the isoelectric point of spike proteins could account for the higher LOD and narrower LDR than found in RSV, as the gate could repel some spike proteins while others are attracted.

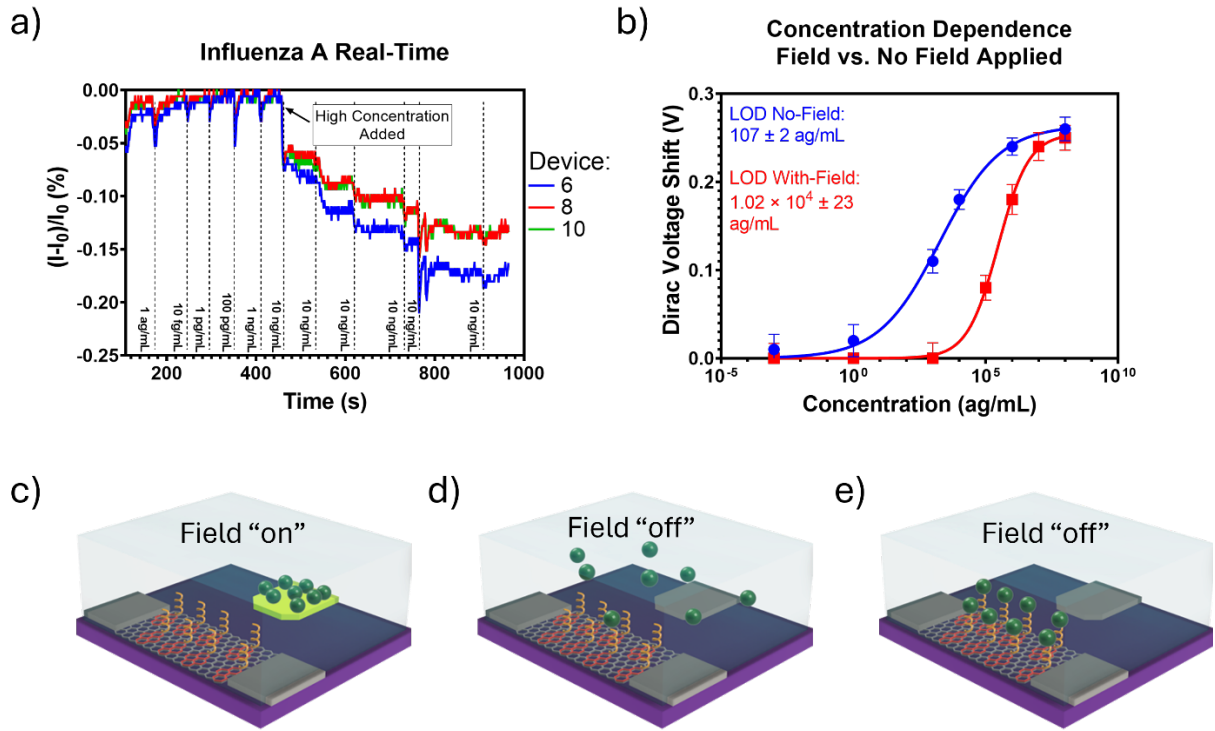

Figure 5. a) Real-time measurement of Influenza A Hemagglutinin. Little to no change in current was noted until high concentrations of the target were added. b) Dirac voltage shift as a function of concentration for two cases: *blue* – no applied electric field from the gate, *red* – gate voltage set to the maximum transconductance value only during hour-long incubations of each concentration. This static field is turned off during Dirac point voltage measurements, and the gate voltage is swept as per normal protocol. The LOD was found to be two orders of magnitude lower in the absence of applied field. c) With the gate "on," the hemagglutinin is pulled away from the aptamers preventing binding. d) With the gate "off," negatively charge hemagglutinin proteins in solution. e) With the gate "off," we have found that one-hour incubation time is sufficient for proteins in solution to bind with aptamers.

To determine whether the applied gate voltage is indeed the source of the difference in our transconductance results, we conducted new Dirac-point protocol measurements. Specifically, the gate voltage was held either at the transconductance peak (active field) or turned off (no applied field) during analyte incubation. The post-incubation Dirac point shifts revealed a two orders of magnitude difference in LOD between the two conditions: 107 ± 2 ag/mL with no applied field versus $1.02 \times 10^4 \pm 23$ ag/mL under continuous gating. The applied field, therefore, hindered analyte capture, reducing binding efficiency and compressing the dynamic range from 5.0 orders of magnitude (no field) to 2.6 orders of magnitude (with field) (Figure 5b). These findings underscore a critical and previously unexplored interplay among analyte charge polarity, gate bias, pH, and aptamer orientation in liquid-gated GFET sensors.

# 4. Conclusion

This study directly addresses a central question in graphene biosensing: how to identify the most effective electrical readout strategy for detecting biomolecular interactions in complex, real-world environments. By systematically comparing Dirac point tracking and transconductance monitoring in aptamer-functionalized GFETs, we reveal how the choice of readout fundamentally determines sensing performance under high ionic strength and heterogeneous sample conditions typical of complex media.

Our results demonstrate that Dirac point tracking provides a broad dynamic range and quantitative precision in controlled settings. However, improvement of one to two orders of magnitude in sensitivity via transconductance-based readouts is achievable, though at the cost of lower reliability, reliance on nearly neutral or positively charged targets, and a much lower dynamic range. Indeed, our work shows that the same device can be employed in different modalities to optimize for distinct goals (i.e., dynamic range versus LOD). This also shows that reported LODs and dynamic ranges must be considered in the context of the electronic readout scheme employed. Furthermore, from a device design perspective, it is necessary to tailor gating polarity and magnitude to the target biomolecule's net charge.

Our work also offers insight into a long-standing mystery in transconductance measurements: the origin of the large initial dips in current when fluid is added. Specifically, our adoption of a co-planar side-gate configuration suppresses these dips and enhances device stability by minimizing electrostatic and hydrodynamic perturbations that commonly limit top-gated designs. Together, these improvements reduce detection limits for viral proteins to the attogram-per-milliliter regime (Table 2), highlighting the promise of transconductance-based GFETs for ultra-trace biosensing.

In the context of wastewater surveillance, these insights establish a practical framework for field-deployable pathogen monitoring: Dirac point tracking for quantitative, longitudinal measurements across wide concentration ranges, and transconductance analysis for early-stage detection when viral signals are weakest. In wastewater matrices, where ionic strength and surface adsorption events fluctuate, the Dirac point may shift ambiguously or remain masked by background potential drifts. However, the derivative nature of transconductance emphasizes differential changes in channel mobility and carrier modulation efficiency, filtering out low-frequency drift components and enhancing sensitivity to molecular interactions. Transconductance reflects the slope of the transfer curve rather than its absolute position, offering resilience against environmental variability.

**Table 2.** Summary of GFET biosensing performance across viral targets using Dirac point and transconductance readouts.

| Target Analyte | Gate Configuration | Readout Type | Limit of Detection (LOD) | Linear Dynamic Range ($\log_{10}$) | Key Observations / Notes |
|---|---|---|---|---|---|
| **SARS-CoV-2 Spike Protein** | Top Gate | Transconductance | 275 ± 10.3 ag/mL | 8.6 (n = 0.22) | Strong transient current dips from ion redistribution; less stable gating; baseline drift observed. |
| | Side Gate | Transconductance | 16 ± 0.25 ag/mL | 3.8 (n = 0.50) | Stable baseline, improved reproducibility; higher SNR due to reduced ionic disturbance and better electrostatic coupling. |
| | Side Gate | Dirac Point (previous work) | $1.4 \times 10^5 \pm 1.2 \times 10^4$ ag/mL | 4.6 (n = 0.42) | Lower sensitivity; broad quantifiable range. |
| **Respiratory Syncytial Virus (RSV) Glycoprotein** | Side Gate | Transconductance | 10.8 ± 1.2 ag/mL | 4.5 (n = 0.42) | Rapid stabilization, minimal post-injection artifacts; improved sensitivity over Dirac tracking. |
| | Side Gate | Dirac Point (previous work) | $1.8 \times 10^5 \pm 2.3 \times 10^4$ ag/mL | 4.9 (n = 0.39) | Broader dynamic range but higher noise in complex matrices. |
| **Influenza A Hemagglutinin (HA)** | Side Gate | Dirac Point (field applied) | $1.0 \times 10^3 \pm 23$ ag/mL | 2.6 (n = 0.75) | Positive gate bias repelled negatively charged HA; reduced binding efficiency. |
| | Side Gate | Dirac Point (no applied field) | 107 ± 2 ag/mL | 5.0 (n = 0.38) | Highest sensitivity achieved when field is off; demonstrates analyte charge–bias interaction. |

# Author Information

## Corresponding Author

Kenneth S. Burch – Department of Physics, Boston College, Chestnut Hill, Massachusetts 02467, United States; orcid.org/0000-0002-7541-0245; Email: [ks.burch@bc.edu](ks.burch@bc.edu)

## Authors

Michael Geiwitz – Department of Physics, Boston College, Chestnut Hill, MA 02467, United States; orcid.org/0009-0000-7197-9381

Owen Rivers Page – Department of Biology, Boston College, Chestnut Hill, Massachusetts 02467, United States; orcid.org/0000-0003-4072-8509

Tio Marello – Department of Physics, Boston College, Chestnut Hill, MA 02467, United States

Marina E. Nichols – Department of Physics, Boston College, Chestnut Hill, MA 02467, United States; orcid.org/0009-0006-4355-809X

Catherine Hoar – Department of Engineering, Boston College, Chestnut Hill, MA 02467, United States; orcid.org/0000-0001-6287-9186

Deji Akinwande – Department of Biomedical Engineering, The University of Texas at Austin, Austin, Texas 78758; orcid.org/0000-0001-7133-5586

Michelle M. Meyer – Department of Biology, Boston College, Chestnut Hill, Massachusetts 02467, United States; orcid.org/0000-0001-7014-9271

## CRediT authorship contribution statement

K.S.B. and M.G. conceived the project and designed the experiments. M.G., M.E.N. and T.M. fabricated, functionalized, and tested all GFET platforms. M.G., O.R.P, M.M. and K.S.B. selected all aptamers and analyzed the data. O.R.P. pre-linked all aptamers with PBASE linker molecules. T.M. and M.E.N. aided in the design and placement of acrylic wells and the handheld sensing platform. M.M. supervised aptamer pre-linking and sample preparation. C.H., D.A. and K.S.B. aided in data analysis. M.G. and K.S.B. wrote the manuscript with the contribution of all co-authors.

## Data Availability

All data is available on request.

## Acknowledgements

The work of M.E.N. and T.M. was supported by National Science Foundation, Award No. DMR-2310895. M.M.M., M.G., K.S.B. and O.R.P. are grateful for support from the Boston College Schiller Institute for Integrated Science and Society Exploratory Collaborative Scholarship grant. M.G. would like to thank the Boston College Cleanroom & Nanofabrication Facility's staff, Stephen Shepard (retired) and Dr. Chris Gunderson. M.G. would also like to thank Mark Bukowski from the Veolia wastewater treatment plant for wastewater samples. D.A. acknowledges a research grant from the Defense Threat Reduction Agency (DTRA), HDTRA1-24-1-0020.

## Supporting Information

The supporting information is available free of charge.

Bar graph showing initial Dirac point variability for various well configurations (Figure S1); Clover well images showing functionalization technique (Figure S2); Aptamer pre-linking process (Supplemental S3); Dirac point tracking curves with no applied gate voltage during target incubation and with active gate voltage set to maximum transconductance point (Figure S4).

## Conflict of Interest Disclosure

The authors declare that they have no known competing financial interests or personal relationships that could have appeared to influence the work reported in this paper.